\documentclass[paper,11pt]{JHEP}
\usepackage[centertags]{amsmath}
\usepackage{cite}
\usepackage{amsfonts} \usepackage{amssymb} \usepackage{amsthm}
\allowdisplaybreaks[1]
\usepackage{graphicx}
\usepackage{slashed}
\usepackage{latexsym}
\usepackage{bm}
\usepackage[utf8]{inputenc}
\usepackage{amsmath,mathtools}
\numberwithin{equation}{section}

\DeclareMathOperator{\arcsinh}{arcsinh}
\newcommand{\tr}{\mathrm{tr}\,}

\newcommand{\dd}{\mathrm{d}}

\newcommand{\trK}{\mathrm{tr_{K}}\,}

\usepackage{graphics,psfrag}
\usepackage{amsthm,amssymb,epsfig,amsmath,euscript,array,cite}
\usepackage{cancel}
\usepackage{graphicx}
\usepackage{amscd}
\usepackage{slashed}

\newcommand{\be}{\begin{equation}}
\newcommand{\ee}{\end{equation}}
\def\bea{\begin{eqnarray}}
\def\eea{\end{eqnarray}}

\def\nn{\nonumber}

\title{ Wilson loop and its correlators in the limit of large coupling constant}
\author{E.Sysoeva\\
Dipartimento di Fisica\\
	Universit\`a di Roma Tor Vergata\\
	I.N.F.N - sezione di Roma Tor Vergata \\
	Via della Ricerca Scientifica, I-00133 Roma, Italy\\
	\email{{\it Email:} ekaterina.sysoeva@roma2.infn.it}
}

\abstract{In this paper we study Wilson loops in various representations for finite and large values of the color gauge group for supersymmetric ${\cal N}=4$ gauge theories. We also compute correlators of Wilson loops in different representations and perform a check with the dual gravitational theory.}


\preprint{ROM2F/2017/05}

\begin{document}

\section{Introduction}
Supersymmetric Wilson loops as well as their correlators both with chiral primary operators and with other Wilson loops are remarkable observables of the supersymmetric Yang-Mills gauge theories (SuSy YM) and provide stringent tests of AdS/CFT correspondence. 
Due to the fact that the propagator is constant on a circle the computation on the CFT side can be performed with a matrix model as it was demonstrated with the help of localization \cite{Pestun}.
\par As discussed in a number of papers \cite{NickH,Okuda:2008px} in the case of large coupling constant the computation of a Wilson loop's vacuum expectation value in the framework of a matrix model can be significantly simplified. In the present paper we consider the very same limit of large coupling constant and perform some matrix model calculations for the vacuum expectation value of the Wilson loop in different representations, for its correlators with another Wilson loop and with chiral operators both for finite number of colours $N$ of the gauge theory and for large $N$. Furthermore, we compare a correlator between two $\frac{1}{2}$-BPS Wilson loops, one of which is in the fundamental representation of the gauge group and the other in a representation associated with a Young tableau with several long lines, with the corresponding quantity on the AdS side and find perfect agreement.
\par The paper is organized as follows: in Section 2 we calculate a vacuum expectation value of a Wilson loop in a general representation and perform calculation for finite number of colours $N$. We proceed with considering  the large $N$ limit of the previous case. In Section 3 we turn to the correlator of a symmetric Wilson loop with primary chiral operators, again both for finite and large $N$. Finally in Section 4 we study the correlator of the two Wilson loops discussed in the previous sections both from the point of view of the matrix model and on the AdS side.

\section{Wilson loops in arbitrary  representations}

We consider a $\frac{1}{2}$-BPS circular Wilson loops in ${\cal N}=4$ super Young-Mills theory with gauge group $U(N)$. The vacuum expectation value of the WL defined as
\begin{equation}
W_{\bf R} = \frac{1}{N}\left<\tr_{\bf R} \, e^{\mathcal{C}} \right>_{vev}= \frac{1}{N}\left< \tr_{\bf R} P \, \exp \left[ \oint_{\cal C } ds \left(i A_{\mu} \dot{x^{\mu}} + \vec{n}\cdot \vec{\Phi}(\dot{x}) \right) \right] \right>_{vev}
\end{equation}
due to localization can be found as a $U(N)$ matrix model integral \cite{Pestun}
\begin{equation}
W_{\bf R} = \frac{1}{N} \left<  \tr_{\bf R}\, e^{a}  \right>.
\end{equation}
The averages in the matrix model are defined as
\begin{equation}
\left< f(a) \right> = \frac{1}{Z}\int  d a \, \Delta(a) \,e^{-   \sum_u {a_u^2}   }\, \, f(g \, a) \, ,
\end{equation}
where $g=\sqrt{\frac{\lambda}{2N}}$ and 
\begin{equation}
Z=\int  d a \, \Delta(a) \,e^{-   \sum_u {a_u^2}   }\,
\end{equation}
with $da=\prod_{u=0}^{N-1} da_u$ being the Lebesgue measure in the space of eigenvalues of the matrix $a$ in the fundamental representation (absorbing numerical coefficient irrelevant for the calculation of averages) and $\Delta(a)^{\frac{1}{2}}$ being the Vandermonde determinant
\begin{equation}
\Delta(a)^{\frac{1}{2}}= \prod _{u<v=0}^{N-1} \left(a_u-a_v\right) \, .
\end{equation}
A representation ${\bf R}$ of the $U(N)$ group is specified by the  Dynking labels $\lambda=(\lambda_0,\lambda_1,\ldots \lambda_{N-2})$ and central charge $Q$, or equivalently by a Young tableau with rows of length $K_u$ given by
 \be \label{lenghts}
K_u=\sum_{j=u}^{N-2} \lambda_j +\frac{Q-\sum_{j=0}^{N-2} \lambda_j}{N} \, ,  \quad\quad  u=0,\ldots N-1   \,\, .
 \ee  
Let us introduce the orthonormal basis $\{ e_i \}$ with 
 $e_i \in \mathbb{R}^N$ and write the $U(N)$ simple roots as $\alpha_i=e_i-e_{i+1}$ for $i=0,\ldots N-2$. The character of a representation is given by the Weyl formula
  \be
\chi = \tr_{\bf R} e^{  g\, a }=\sum_{\alpha \in {\bf R} }  e^{g\,  a\cdot \alpha }=   \frac{{\rm det}_{u,v}     e^{ g\, a_u (K_v +N-v )  }   }{   {\rm det}_{u,v}     e^{ g\,  a_u (N-v )  }  } 
  \ee
  with the sum running over the set of weights $\{ \alpha  \}$ defining the representation ${\bf R}$.   The determinant in the numerator can be written as
  \be
   {\rm det}_{u,v}     e^{ g\,  a_u (K_v +N-v )  }  = \sum_{\sigma\in S_N}  (-)^{\sigma}\,   \prod_{u=0}^{N-1}  e^{g\,  a_{\sigma_u} (K_u +N-u )  }  \, ,
  \ee
   while that in the denominator can be explicitly computed and written in the form
  \be \label{denominator}
  {\rm det}_{u,v}     e^{ g\, a_u (N-v )  }  =\prod_{u<v}  \left(   e^{  g\,  a_u  } -   e^{  g\, a_v  }    \right)   \, .
  \ee
It can be noted that (\ref{denominator}) is invariant under permutation up to a sign of the permutation, so we can replace $a_u,a_v$ by $a_{\sigma_u},a_{\sigma_v}$ and get rid of $(-)^{\sigma}$ in the nominator. 
  \be  \label{WL}
W_{\bf R} = \frac{1}{N}\frac{1}{Z}\int  d a \, \Delta(a) \, \sum_{\sigma\in S_N}   
\frac{ \prod_{u=0}^{N-1}  e^{-  a_{u}^2+ g\, a_{\sigma_u} K_u    }    }{ \prod_{u<v}  \left(  1 -   e^{ g\, ( a_{\sigma_v}-a_{\sigma_u})  }    \right) } 
\ee
We also notice that the integrals over the eigenvalues are equal for each one of the N! permutations, so we can choose one permutation (for example the trivial one) and rewrite (\ref{WL}) as follows
\be \label{WLtemp}
W_{\bf R} = \frac{1}{N}\frac{1}{Z}\int  d a \, \Delta(a) \, N!
\frac{ \prod_{u=0}^{N-1}  e^{-  a_{u}^2+ g\, a_u K_u    }    }{ \prod_{u<v}  \left(  1 -   e^{ g\, ( a_v-a_u)  }    \right) } \, .
\ee
In the strong coupling limit $g \gg 1$ the product in the denominator of (\ref{WLtemp}) is equal to $1$ in the region $\Omega \subset \mathbb{R}^n$ such that $a_{v} <a_{u} $ for all $v>u, \, u=0, \, ... \, N-1$  and diverges otherwise, so only in $\Omega$ the integrand is not suppressed exponentially by the denominator. Hence we actually have
\be \label{WLtemp2}
W_{\bf R} = \frac{1}{N}\frac{1}{Z}\int_{\Omega}  d a \, \Delta(a) \, N!
\prod_{u=0}^{N-1}  e^{-  a_{u}^2+ g\, a_u K_u    } \, .
\ee
Thinking of a representation associated with a Young tableau with $g+1$ groups consisting of $n_i$ rows of the same length (including rows of zero length) one can notice that the integrand of (\ref{WLtemp2}) stays the same under the permutation $u \rightarrow \sigma_u$ such that $K_u=K_{\sigma_u}$. It implies that the integral (\ref{WLtemp2}) over $\Omega$ can be replaced with the integral over the union of the images of $\Omega$ under all such permutations (let us denote it as $\tilde{\Omega}$) divided by the number of the permutations $n_1 ! \, n_2 ! \ldots n_{g+1}!$.
Clearly $\tilde{\Omega}$ includes all regions where $a_v<a_u$ whenever $K_v<K_u$. We then get
\be \label{WLtemp3}
W_{\bf R} =  \frac{c_{\bf R}}{N \, Z}\int_{\tilde{\Omega}}  d a \, \Delta(a) \,
 \prod_{u=0}^{N-1}  e^{-  a_{u}^2+ g\, a_u K_u } 
\ee
with
\begin{equation}
  c_{\bf R}=\frac{N! }{ n_1 ! \, n_2 ! \ldots n_{g+1}!  } \, .
\end{equation}
Finally, we can extend the integral (\ref{WLtemp3}) back to $\mathbb{R}^n$ from $\tilde{\Omega}$ since the integrand is suppressed exponentially everywhere except $\tilde{\Omega}$.
We finally get
 \footnote{
  In the case of the antisymmetric representation associated with a Young tableau with one column of the length $l$ one finds
    \be
W_{\bf R} =  \frac{1}{Z \, N}  \frac{N! }{(N-l)!l!}\int  d a \, \Delta(a) \,e^{ -\sum_u  { a_u^2} + \sum_{i=0}^{l-1} g\, a_i \,     }\,  
\ee
   We notice that in this case the result is exact.
   \par In particular for the fundamental representation one gets
  \be
W_{\bf R} =   \frac{ 1 }{Z}\int  d a \, \Delta(a) \,e^{ -\sum_u { a_u^2 } + g\, a_0     }\,\, .
\ee
}
 \be  \label{WLleading}
W_{\bf R} \approx    \frac{c_{\bf R} }{ZN}\int  d a \, \Delta(a) \,e^{ \sum_u  \left( -{ a_u^2 } + g\, a_u \, l_u \right)    }\, . 
\ee
The integral (\ref{WLleading}) can be explicitly computed for finite $N$. 
To  this purpose we write the Vandermonde determinant in the
\be
\Delta(a)^{\frac{1}{2}}= \prod _{u<v=0}^{N-1} \left(a_u-a_v\right)= 2^{-\frac{N(N-1)}{4}}{\rm det} \, (\sqrt{2 }a_u)^v =
\ee
$$
=  2^{-\frac{N(N-1)}{4}}   {\rm det} \, \left( 2^{-\frac{v}{2}} H_v\left(a_u\right) \right)   =  2^{-\frac{N(N-1)}{2}}\sum_{\sigma} (-)^\sigma \prod_u   H_{\sigma_u} \left(a_u\right)
$$
with $H_n(x)$ being the "physicists" Hermite polynomials \footnote{ Physicists Hermite polynomials  are defined such that  $H_n=2^n x^n+\ldots $, \textit{i.e. }
\be
H_n=\{ 1,2x,4x^2-2,8x^3-12\,x,16x^4-48\, x^2+12,\ldots    \}
\ee
}
\be
H_n(x)=e^{x^2} \left( -{\frac{d}{d \, x}}\right)^n e^{-{x^2} } \, .
\ee 
  The Hermite polynomials satisfy the  integral relations  
\bea
&&  \int _{-\infty }^{\infty }dx \, H_m (x) \, H_n(x)\, e^{-x^2 } =\sqrt{\pi }2^n \,n!  \, \delta _{mn}  \, , \nn\\
&& \int _{-\infty }^{\infty }dx \,  H_n (x)^2 \, e^{-x^2+x t} = \sqrt{ \pi } 2^n  n !  \,  {\rm L} _n \left(-\frac{t^2}{2}\right)\, e^{\frac{t^2}{4} } \, ,  \nn\\
&& \int _{-\infty }^{\infty }dx \,H_m (x)\,  H_n (x)\,  \, e^{x^2+x t} = \sqrt{\pi }  m ! \, 2^{m} \, t^{n-m}\,  {\rm L} ^{n-m}_m \left(-\frac{t^2}{2}\right)\, e^{\frac{t^2}{4}}
  \label{identint}
\eea
with
\be
L_n(x) = \frac{1}{n!} \, e^x \, \left( \frac{d}{d \, x}\right)^n \left(  e^{-x} \, x^n \right)
\ee
standing for the Laguerre polynomials and
    \be
L^\alpha_n(x) = {\frac{1}{n!}} \, e^x \, x^{-\alpha} \left( \frac{d}{d \, x}\right)^n \left(  e^{-x} \, x^{n+\alpha}\right) =\sum_{j=0}^n \left(
\begin{array}{c}
  n+\alpha   \\
  n-j  \\   
\end{array}
\right)  \frac{ (-x)^j}{ j!}
\ee
standing for the generalized ones. Using these  integral relations   one can compute  the integrals as follows
 \bea \label{ZK}
  Z(\vec K)  &= &    c_{\bf R}  \, \int  d a \, \Delta(a) \,e^{ \sum_u  \left( -{ a_u^2 } + g\, a_u \, K_u \right)    } \\
&=& c_{\bf R} \,  2^{-\frac{N(N-1)}{2}} \, \sum _{\sigma ,\sigma '}  (-)^{ \sigma +\sigma ' }  \int _{-\infty }^{\infty  }\prod _u \left(da_u  \,e^{-a_u^2} 
H_{\sigma _u}\left[a_u  \right] H_{\sigma '_u}\left[a_u \right]    e^{g\, K_u  a_u }    \right)  \nn\\
&=&   \pi^{\frac{N}{2}} \,  c_{\bf R}  \, \sum _{\sigma,\sigma' }  (-)^{ \sigma +\sigma ' }
\prod _u \left[  \ \sigma_u ! \,  \left({ g\, K_u  }\right)^{\sigma_u'-\sigma_u }\,    e^{ \frac{g^2\, K_u^2}{4}}     L_{\sigma _u}^{\sigma_u'-\sigma_u} \left( - \frac{ g^2\, K_u^2    }{ 2 } \right)     \right]  \nn\\
&=&   \pi ^{\frac{N}{2}} \,  c_{\bf R}  \, G(N+1) \, e^{ 2N \sum_u y_u^2 }  \sum _{\sigma,\sigma' }  (-)^{ \sigma +\sigma ' }
\prod _u \left[    K_u^{\sigma_u'-\sigma_u }\,       L_{\sigma _u}^{\sigma_u'-\sigma_u} \left( - 4 N y_u^2 \right)    \right]  \nn
\eea
with $G(N+1)=\prod_{u=0}^{N-1} u!$ and $y_u= \frac{K_u \sqrt{\lambda}}{4 N}$. In the following subsections we specify to some simple cases.

  \subsection{ The completely symmetric representation}
  
Let us consider the K-symmetric representation, characterised by a Young tableau with a single row of length $K$, \textit{i.e.} $K_u=K\, \delta_{u0}$.   
Formula (\ref{ZK}) reduces to
$$
 Z(K) =  c_{\bf R}  \, \int _{-\infty }^{\infty  }d^Na   \prod _{u<v=0}^{N-1} \left(a_u-a_v\right){}^2 e^{-   \sum_u {a_u^2 }  +g\,  K a_0} = 
$$
\begin{equation}
= \pi^{\frac{N}{2}} \,  c_{\bf R}  \,     \, G(N+1) \, (N-1)!\, e^{ 2 N y^2 }  \sum _{u }  
   L_{u} \left( -4 N y^2 \right)
 \end{equation}
leading to
\begin{equation} \label{WLsym}
W_K =\frac{1}{N} \frac{Z(K) }{ Z(0) }= \frac{1}{N}\,e^{2 N y^2 }  \, \sum _{u=0 }^{N-1} { \rm L}_{u}\left( -  4 N y^2\right) = \frac{1}{N}\,      e^{ 2 N y^2 }  \,
{ \rm L}^1_{N-1} ( -  4 N y^2) \, .
\end{equation}

\subsubsection{Large N limit}

There are various limits one can consider. 

\begin{itemize}

\item{Large $N$ keeping $K$ finite:

In this limit one can write
\be \label{WLKplan}
W_{\bf R} =\frac{e^{2 N y^2}}{N} 
   \sum_{u=0}^{N-1}    
\left(
\begin{array}{c}
  N   \\
 u+1  \\   
\end{array}
\right)
   \frac{\left(  4 N y^2  \right)^u}{u!}  \approx  
   \sum_{u=0}^{\infty}    
  \frac{ \left(    K^2 \lambda   \right)^u}{4^u \, u! \, (u+1)!}  = \frac{2       I_1 \left(     K \sqrt{\lambda}     \right)   }{K \sqrt{\lambda} }  \, ,
\ee
where we used the definition of the Bessel function $I_{n}(x)=\sum_{u=0}^\infty \frac{(x/2)^{2u+n}}{ u!(n+u)!}$.  For large $\lambda$ one finds
\be
W_{\bf R} \approx  e^{ K \sqrt{\lambda} } \, . 
\ee 

}

\item{ $N$ and $K$ tend to infinity with $K/N$ finite. In this limit one can use the large $N$ asymptotic formula of the Laguerre polynomials \cite{Laguerre}
\begin{equation} \label{WKlargeN}
W_{K}=\frac{1}{N} L_{N-1}^1\left(-4Ny^2\right) e^{2Ny^2}
=\frac{1}{N}\frac{\alpha_0(y)}{2F(y)}  
 I_{1}\left(4 N F(y)\right)
\end{equation}
with
$$
\alpha_0(y)=\left( \frac{F(y)}{y} \right)^{2}\left( \frac{y^2}{1+y^2} \right)^{\frac{1}{4}} \frac{1}{\sqrt{F(y)}} \,,
$$
\begin{equation}
F(y)=\frac{1}{2}\left( \arcsinh(y)+ y\sqrt{1+y^2}\right) \, .
\end{equation}
So the WL in this limit is
 \be
W_{K}  =  \frac{1}{N} \, { \rm L}^1_{N-1}\left( - 4 N y^2  \right)\, e^{ 2 N y^2 } \sim    {\rm exp}  
 \left\{  4 N  F(y)  \right\} \, .
\ee

 }
 
 \end{itemize}
The exact expression for the symmetric WL (\ref{WLsym}) can be obtained with the help of diagrams in the frame of the matrix model. One would expect that the large $N$ limit can be restored by only planar diagrams, however, it can be checked that for any $K$ the planar diagrams sum to (\ref{WLKplan}) and never give (\ref{WKlargeN}), \textit{i.e.} the planar diagram approach cannot be applied for the case of finite $K/N$.

\subsection{Two-row Young tableau}

Let us consider now a representation defined by a Young tableau made of two rows of lengths $\vec{K}=\{K_0, K_1 \}$. We write
\begin{equation} \label{WL2l}
W_{\{K_0, K_1 \}}=   \frac{1}{N}\frac{Z(\vec K)}{Z(\vec 0) }
\end{equation}
with
\bea 
 Z(\vec K) &=& c_{\bf R}  \,  \int d^N a \, \Delta(a)\, e^{- \sum_u {a_u^2}+g\,( K_0 a_0+K_1 a_1)} \\
 &=&   \pi^{\frac{N}{2}} \,  c_{\bf R}  \, G(N+1) \, e^{ 2 N (y_0^2+y_1^2)}  \sum _{\sigma,\sigma' }  (-)^{ \sigma +\sigma ' }
\prod _u \left[    \left(K_u\right)^{\sigma_u'-\sigma_u }\,       L_{\sigma _u}^{\sigma_u'-\sigma_u} \left( - 4 N y_u^2  \right)    \right] \nn\\
 &=& \pi^{\frac{N}{2}} \,  c_{\bf R}  \, G(N+1) \, (N-2)!\, e^{ 2 N (y_0^2+y_1^2)} \cdot  \nn \\
 &\cdot&
 \sum _{i,j=0 }^{N-1} 
  \left[L_{j} \left( - 4 N y_0^2  \right) \, L_{j}\left( - 4 N y_1^2  \right)  -    \left(\frac{  y_0}{y_1 } \right)^{i-j }\,  L_{j}^{i-j} \left( -  4N y_0^2 \right) \, L_{i}^{j-i} \left( - 4 N y_1^2\right) \right]   \, .\nn
  \eea
 In this case
 $$
 c_{\bf R}(\vec K)=\frac{N!}{(N-2)!}\left(1-\frac{1}{2}\delta_{y_0,y_1}\right) \, ,
 $$
 so one finds for the Wilson loop
\bea \label{WL2rows}
W_{\{K_0, K_1 \}}&=&\frac{1}{N}  \left(1-\frac{1}{2}\delta_{y_0,y_1} \right) \, e^{2 N (y_0^2+y_1^2) }  \nn \\
&\cdot&
\sum _{i,j=0 }^{N-1} 
  \left[L_{j} \left( - 4 N y_0^2 \right) \, L_{j}\left( -  4 N y_1^2  \right)  -    \left(\frac{  y_0 }{ y_1  }\right)^{i-j }\,       L_{j}^{i-j} \left( -  4 N y_0^2     \right) \, L_{i}^{j-i} \left( -  4 N y_1^2 \right)    \right]  \nn  \\ 
&=&  \frac{1}{N} \left(1-\frac{1}{2}\delta_{y_0,y_1} \right) \, e^{2 N (y_0^2+y_1^2) } \\ \nn
&\cdot& \left[  L_{N-1}^1\left( -  4 N y_0^2    \right) L_{N-1}^1\left( - 4 N y_1^2   \right)  -   \sum _{i,j=0 }^{N-1} \left(\frac{ y_0}{y_1  }\right)^{i-j }\,       L_{j}^{i-j} \left( -4 N y_0^2 \right) \, L_{i}^{j-i} \left( - 4 N y_1^2 \right)  \right]  
\eea

\subsubsection{Large $N$, finite $K_u/N$ limit}

Although it is difficult to extract the large $N$ limit of a WL in the representation associated with a Young tableau with two rows directly from (\ref{WL2rows}), we will need this result later in Section 4. We then use an alternative method to perform such limit for the case of a Young tableau with long rows (the lengths of the rows $K_u$ are of order $N$).
\par The WL in the K-symmetric representation can be written as an integral over $N\times N$ hermitian matrices
\begin{equation}
W_K= \frac{1}{N} \frac{\int d[a] \trK e^a e^{-\frac{2 N}{\lambda} \tr a^2}}{\int d[a] e^{-\frac{2N}{\lambda} \tr a^2}}=\frac{1}{N}\int d\mu[v] \trK e^{v\frac{\sqrt{\lambda}}{2\sqrt{N}}},
\end{equation}
where $d[a]$ is the Lebesgue measure in the space of hermitian $N\times N$ matrices, $v=\frac{\sqrt{N}}{\sqrt{\lambda}}a$ and $d\mu[v]$ is the Gaussian measure in the space of $N\times N$ hermitian matrices 
$$d\mu[v]= 2^{\frac{N^2-N}{2}} \left( 2\pi\right)^{-\frac{N^2}{2}}e^{-\frac{1}{2}tr v^2} d[v], \,\, \int d\mu[v]=1 \, .$$
As it was shown in \cite{Sys} the WL in the limit of large coupling constant $\lambda$ can be rewritten as an integral over $N-1 \times N-1$ hermitian matrices
\begin{equation} \label{WLdoubleint}
W_K=\frac{1}{N!} \int d\mu[v_0]e^{\frac{K v_0}{2}\frac{\sqrt{\lambda}}{\sqrt{N}}}v_0^{2(N-1)}\int d\mu[\tilde{v}]e^{2\, \tr\left( \ln(1- \frac{\tilde{v}}{v_0})\right)},
\end{equation}
where $d\mu[\tilde{v}]$ is the Gaussian measure in the space of $N-1 \times N-1$ hermitian matrices and $d\mu[v_i]$ is the Gaussian measure in the space of real eigenvalues
$$
d\mu[v_i]=\frac{1}{\sqrt{2\pi}}e^{-\frac{v_i^2}{2}}dv_i \, .
$$
Assuming now, that the length of the row $K$ is of order of $N$ one can compute the integral with measure $d\mu[v_0]$ in (\ref{WLdoubleint}) using standard perturbation theory for large $N$ and the other integral with the saddle point method. Taking only the leading order in the large $N$ expansion one will find then that at the saddle point $v_0=2 \sqrt{N} \sqrt{1+y^2}$ (\ref{WLdoubleint}) gives the correct exponential behavior
\begin{equation}
W_K \sim e^{4N F(y)},
\end{equation}
Taking the leading and the first subleading orders in large $N$ expansion one will restore also the correct pre-exponential factor \cite{Sys}
\begin{equation}
W_K \approx \frac{1}{\sqrt{2\pi N}}\left( \frac{y^2}{1+y^2} \right)^{\frac{1}{4}}\frac{1}{N\left(2y\right)^2}e^{-N}e^{4N F(y)} \cdot \left(1+ \mathcal{O}\left(\frac{1}{N}\right) \right).
\end{equation}
Let us now use the same method to find the exponential behavior of a WL in the representation defined by a Young tableau with two rows with lengths $K_0, \, K_1 \sim N$. Acting in the same way as in \cite{Sys} one will find that
\begin{equation}
W_{\vec{K}}=\frac{1}{N!(N-1)!} \int d\mu[v_0]d\mu[v_1]e^{\frac{K_0 v_0}{2}\frac{\sqrt{\lambda}}{\sqrt{N}}}e^{\frac{K_1 v_1}{2}\frac{\sqrt{\lambda}}{\sqrt{N}}} \cdot 
\end{equation}
$$
\cdot \, v_0^{2(N-2)} v_1^{2(N-2)} (v_0-v_1)^2 \int d\mu[\tilde{v}]e^{2\, \tr\left( \ln(1- \frac{\tilde{v}}{v_0})\right)}e^{2\, \tr\left( \ln(1- \frac{\tilde{v}}{v_1})\right)} \, ,
$$
where $d\mu[\tilde{v}]$ is the Gaussian measure in the space of $N-2 \times N-2$ hermitian matrices. If $K_0 \neq K_1$ one will see immediately from the previous case that the saddle points are $v_0=2\sqrt{N}\sqrt{1+y_0^2}, \, v_1=2\sqrt{N}\sqrt{1+y_1^2}$ and in the leading order of the large $N$ expansion
\begin{equation} \label{expbeh2}
W_{\vec{K}} \sim \prod_{u=0}^{1} e^{4N F(y_u)}.
\end{equation}
If now $K_0 = K_1$ one has to be more careful with the factor $(v_0-v_1)^2$ and to find the saddle points with the corrections
$$
v_0=2\sqrt{N}\sqrt{1+y^2} + a, \, v_1=2\sqrt{N}\sqrt{1+y^2} - a, \,
a^2=\frac{y}{\sqrt{1+y^2}}
$$
and hence only the pre-exponential factor is changed, but the exponential behavior (\ref{expbeh2}) stays the same.
\par (\ref{expbeh2}) numerically coincides with the large $N$ approximation of (\ref{WL2rows}).
\par Similarly for a WL in the representation associated with a Young tableau with several lines (number of lines $n \ll N$) with lengths $K_u \sim N$ one gets
\begin{equation} \label{WLsevrows}
W_{\vec{K}} \sim \prod_{u=0}^{n-1} e^{4N F(y_u)}.
\end{equation}

\subsection{ Examples of the representations of U(3) gauge group}

Let us show on some simple examples that the leading order of (\ref{WL}) for U(3)  group
\be
W_{\bf R} = \frac{1}{3}\frac{1}{Z}\int  d a_0 d a_1 d a_2 \, (a_0-a_1)^2 (a_0-a_2)^2 (a_1-a_2)^2 e^{-  (a_0^2+a_1^2+a_2^2)} \chi
\ee
with the character of a representation of U(3) gauge group 
$$
\chi=\sum_{\sigma\in S_3}   
\frac{ \prod_{u=0}^{2}  e^{ g\, a_{\sigma_u}(2-u+ l_u  )  }   }{\prod_{u<v}  \left(  e^{g\,a_{\sigma_u}} -   e^{ g\, a_{\sigma_v} }    \right) } 
$$
given by the approximation (\ref{WLleading}) in the large $\lambda$ limit.

 \begin{itemize}
 \item {Symmetric representation defined by a Young tableau $\{2, 0,0\}$}
 
The character in this case is
\begin{equation}
\chi=e^{2g \,a_1}+e^{2g \,a_2}+e^{2g \,a_3}+
\end{equation}
$$
+e^{g \,(a_1+a_2)}+e^{g \,(a_2+a_3)}+e^{g \,(a_1+a_3)} \, .
$$
The first three terms are leading in some regions of the space of eigenvalues (in agreement with $ c_{\bf R}=\frac{3!}{1! 2!}=3$), while the last three terms are never the leading ones.
\par The WL with all terms (\ref{WL}) in this case is
 \begin{equation} \label{temp1}
 W_{\bf R}=e^{\frac{\lambda}{6}}\left(1+\frac{\lambda}{ 3}+\frac{\lambda^2}{54}\right)+e^{\frac{\lambda}{12}}\left(1+\frac{\lambda}{12}+\frac{\lambda^2}{864}\right)
 \end{equation}
 \par and leaving only the leading terms one gets according to (\ref{WLleading})
  \begin{equation} \label{temp2}
 W_{\bf R} \approx e^{\frac{\lambda}{6}}\left(1+\frac{\lambda}{3}+\frac{\lambda^2}{54}\right)
 \end{equation}
 As it can be seen in the $\lambda \gg 1$ limit the difference between (\ref{temp1}) and (\ref{temp2}) is exponentially suppressed.
\par The analytical answer (\ref{WLsym}) is
\begin{equation}
W_{\bf R}=\frac{e^{\frac{\lambda}{6}}}{3}L_2^1\left( -\frac{\lambda}{3}\right)=e^{\frac{\lambda}{6}}\left(1+\frac{\lambda}{3}+\frac{\lambda^2}{54}\right)
\end{equation}

\item {Antisymmetric representation defined by a Young tableau $\{1,1,0\}$}
\par The character of the representation is
\begin{equation}
\chi=e^{g(a_1+a_2)}+e^{g(a_1+a_3)}+e^{g(a_2+a_3)}
\end{equation}
As one can see all three terms (in agreement with $ c_{\bf R}=\frac{3! }{ 2! 1!}=3$) contribute in some regions of the space of eigenvalues.
\par The exact WL (\ref{WL})
 \begin{equation}
 W_{\bf R}=e^{\frac{\lambda}{12}} \left(1+\frac{\lambda}{12}+\frac{\lambda^2}{864}\right)
 \end{equation}
 in this case coincides with the approximate one given by (\ref{WLleading}) 
  \begin{equation}
 W_{\bf R} \approx e^{\frac{\lambda}{12}}\left(1+\frac{\lambda}{12}+\frac{\lambda^2}{864}\right) \, .
 \end{equation}
 The analytical answer (\ref{WL2rows}) is
 \begin{equation}
 W_{\bf R}=\frac{1}{3\cdot 2}e^{\frac{\lambda}{12}} \left( L_2^1\left( -\frac{\lambda}{12}\right) L_2^1\left( -\frac{\lambda}{12}\right)-\sum_{i,j=0}^{2} L_{j}^{i-j} \left(  -\frac{\lambda}{12} \right) \, L_{i}^{j-i} \left(  -\frac{\lambda }{12}  \right) \right)=
 \end{equation}
 $$
 =e^{\frac{\lambda}{12}}\left(1+\frac{\lambda}{12}+\frac{\lambda^2}{864}\right) \, .
 $$
 
 \item {Representation defined by a Young tableau $\{3,2,0\}$}
\par The character of the representation is
\begin{equation}
\chi=e^{g(3a_1+2a_2)}+e^{g(2a_1+3a_2)}+e^{g(3a_3+2a_2)}+
\end{equation}
$$
+e^{g(2a_3+3a_2)}+e^{g(3a_1+2a_3)}+e^{g(2a_1+3a_3)}+
$$
$$
+e^{g(3a_1+a_2+a_3)}+e^{g(a_1+3a_2+a_3)}+e^{g(a_1+a_2+3a_3)}+
$$
$$
+e^{2g(2a_1+2a_2+a_3)}+e^{2g(a_1+2a_2+2a_3)}+e^{2g(2a_1+a_2+2a_3)}
$$
Again one can see that only the first six terms (in agreement with $ c_{\bf R}=\frac{3!}{1! 1! 1!}=6$) are the leading ones in some regions of the space of eigenvalues.
 \par The exact WL (\ref{WL}) in this case is
 \begin{equation}
 W_{\bf R}=e^{\frac{13\lambda}{24}}\left(2+\frac{7\lambda}{6}+\frac{49\lambda^2}{432}+\frac{\lambda^3}{288}\right)+e^{\frac{11\lambda}{24}}\left(1+\frac{\lambda}{3}+\frac{\lambda^2}{54}\right)+e^{\frac{3\lambda}{8}}\left(2+\frac{\lambda}{6}+\frac{\lambda^2}{432}\right)
 \end{equation}
 \par and  (\ref{WLleading}) gives
  \begin{equation}
 W_{\bf R} \approx e^{\frac{13\lambda}{24}}\left(2+\frac{7\lambda}{6}+\frac{49\lambda^2}{432}+\frac{\lambda^3}{288}\right)
 \end{equation}
The difference is again exponentially suppressed in the $g \gg 1$ limit.
\par The analytical answer (\ref{WL2rows}) is
 \begin{equation}
 W_{\bf R}=\frac{1}{3}e^{\frac{13 \lambda }{24}} \left( L_2^1\left( -\frac{3 \lambda}{4}\right) L_2^1\left( -\frac{\lambda}{3}\right)-\sum_{i,j=0}^{2} \left( \frac{3}{2}\right)^{i-j} L_{j}^{i-j} \left(  -\frac{ 3 \lambda }{4} \right) \, L_{i}^{j-i} \left(  -\frac{\lambda} {3}  \right) \right)=
 \end{equation}
 $$
 =e^{\frac{13\lambda}{24}}\left(2+\frac{7\lambda}{6}+\frac{49\lambda^2}{432}+\frac{\lambda^3}{288}\right)
 $$

 \end{itemize}

\section{Correlators of a symmetric Wilson loop and chiral primary operators}

Let us find now a connected correlator between a WL in a symmetric representation and a chiral primary operator. This correlator can be written via matrix model integrals \cite{Fucito}.
\begin{equation}
\left< \trK e^a \, \tr a^n \right>_c = \left<\trK e^a \, \tr a^n \right> - \left< \trK e^a\right> \left< \tr a^n \right>
\end{equation}
where 
\begin{equation}
\left<\trK e^a \, \tr a^n \right> =\frac{1}{Z}\int  d a \, \Delta(a) \,e^{-   \sum_u {a_u^2}   }\, \left( \sum_{i=0}^{N-1} a_i^n \right) \,  e^{  K\, g\, a_0 } \,\, ,
\end{equation}
$$
\left<\tr a^n \right> =\frac{1}{Z}\int  d a \, \Delta(a) \,e^{-   \sum_u {a_u^2}   } \left( \sum_{i=0}^{N-1} a_i^n \right)\, \, .
$$
Both integrals above can be evaluated in the same way as $Z(\vec{K})$ (\ref{ZK}). Having it done, one gets
\begin{equation} \label{corK}
\left< \trK e^a \, \tr a^n \right>_c =\left( \frac{\sqrt{\lambda}}{4N}\right)^n\left(\frac{\dd}{\dd y}\right)^n L_{N-1}^1 \left(-4Ny^2\right) e^{2 N y^2}-
\end{equation}
$$
-\frac{(2-\delta_{s, \,0})}{2^n}\left( \frac{\lambda}{N}\right)^{\frac{n}{2}}e^{2 N y^2}\sum_{s=0}^{n}\sum_{i=0}^{N-1-s}\frac{i! \,(2 \sqrt{N} y)^{s} }{(i+s)!} L_i^{s}\left(-4 N y^2 \right)\frac{\dd^n}{\dd t^n}\left(t^{s}L_{i}^{s}(-t^2)e^{\frac{t^2}{2}}\right)\bigg|_{t=0} \, .
$$
If $n \ll N$ the terms with $i<\frac{n-\delta}{2}$ can be dropped and (\ref{corK}) can be rewritten with hypergeometric functions
\begin{equation} \label{corK2}
\left< \trK e^a \, \tr a^n \right>_c  = \left( \frac{\sqrt{\lambda}}{4N}\right)^n \left(\frac{\dd}{\dd y}\right)^n L_{N-1}^1 \left(-4Ny^2\right) e^{2 N y^2}-
\end{equation}
$$
-\frac{(2-\delta_{s, \,0})}{2^n}\left( \frac{\lambda}{N}\right)^{\frac{n}{2}}e^{2 N y^2}\sum_{s=0}^{n}\frac{n!(2\sqrt{N}y)^{s}}{\epsilon ! \,\rho !}\sum_{i=\epsilon}^{N-1-s}\frac{i! \,  }{(i-\epsilon)!} L_i^{s}\left(-4 N y^2 \right){_2F_1} \left(-\rho,-\epsilon,i+1-\epsilon,\frac{1}{2} \right)
$$
where $\epsilon=\frac{n-s}{2}$, $\rho=\frac{n+s}{2}$.

\subsection{Large N, finite $K/N$  limit}

To calculate (\ref{corK2}) in the first non-vanishing order, for $N \rightarrow \infty$ and finite $K/N$, an integral representation of Laguerre polynomials can be used.
\begin{equation} \label{Lag}
e^{-\frac{x}{2}}L_n^{\alpha}(x)=\frac{1}{2 \pi i} \int_{-\infty}^{0+}e^{-\frac{x}{2}\frac{1+e^{-z}}{1-e^{-z}}}(1-e^{-z})^{-\alpha-1}e^{nz} dz
\end{equation}
Let us consider the second term in (\ref{corK2}) in the limit of large $N$
\begin{equation} \label{termB}
B=\frac{(2-\delta_{s,0})}{2^n}\left( \frac{\lambda}{N}\right)^{\frac{n}{2}}e^{2 N y^2}\sum_{s=0}^{n}\frac{n!(2\sqrt{N}y)^{\delta}}{\epsilon ! \,\rho !}\sum_{i=\epsilon}^{N-1-s}\frac{i! \,  }{(i-\epsilon)!} L_i^{s}\left(-4 N y^2 \right){_2F_1} \left(-\rho,-\epsilon,i+1-\epsilon,\frac{1}{2} \right) .
\end{equation}
Here
\begin{equation}
_2F_1 \left(-\rho,-\epsilon,i+1-\epsilon,\frac{1}{2}  \right)=\epsilon !\rho !\sum_{k=0}^{\epsilon}\frac{\left(i-\epsilon\right)!}{2^k k! \left(\epsilon -k \right)!\left(\rho-k \right)! \left(i-\epsilon + k \right)!}
\end{equation}
Using (\ref{Lag}) one gets a sum over $i$ inside $B$
\begin{equation} \label{midsum}
\sum_{i=\epsilon}^{N-1-s} i! \, e^{z i} \sum_{k=0}^{\epsilon}\frac{1}{2^k k! \left(\epsilon -k \right)!\left(\rho-k \right)! \left(i-\epsilon + k \right)!} \approx \sum_{k=0}^{\epsilon}\frac{1}{2^k k! \left(\epsilon-k \right)!\left(\rho-k \right)!}\sum_{i=\epsilon}^{N-1-s} i^{\epsilon-k}e^{z i} 
\end{equation}
We left only the highest power of $i$ since this term gives the highest power of $N$.
\par (\ref{midsum}) can be further transformed as following
$$
\sum_{k=0}^{\epsilon}\frac{1}{2^k k! \left(\epsilon-k \right)!\left(\rho-k \right)!}\frac{d^{\epsilon-k}}{dz^{\epsilon-k}}\sum_{i=\epsilon}^{N-1-s} e^{z i} =\sum_{k=0}^{\epsilon}\frac{1}{2^k k! \left(\epsilon-k \right)!\left(\rho-k \right)!}\frac{d^{\epsilon-k}}{dz^{\epsilon-k}} \frac{e^{(N-s)z}-e^{\epsilon \, z}}{e^z-1} \approx
$$
$$
\approx \sum_{k=0}^{\epsilon}\frac{1}{2^k k! \left(\epsilon-k \right)!\left(\rho-k \right)!} \sum_{l=0}^{\epsilon-k}(-1)^l C_{\epsilon-k}^ll! \frac{e^{l \, z}}{(e^z-1)^{l+1}}  N^{\epsilon-k-l} e^{(N-s)z}=
$$
$$
=\sum_{l=0}^{\epsilon}(-1)^l \frac{e^{l \, z}}{(e^z-1)^{l+1}} N^{\epsilon-l} e^{(N-s)z}\sum_{k=0}^{\epsilon-l}\frac{N^{-k}}{2^k k!\left(\rho-k \right)!\left(\epsilon-k-l \right)!} 
\approx 
$$
\begin{equation}
\approx  \frac{N^{\epsilon}}{\rho !}\sum_{l=0}^{\epsilon}(-1)^l\frac{e^{(N-s-1)z}}{(1-e^{-z})^{l+1}}N^{-l}\frac{1}{\left(\epsilon-l \right)!} \, .
\end{equation}
Substituting the result back into (\ref{termB}) and using (\ref{Lag}) again one finally gets the second term of  (\ref{corK2}) in the form
\begin{equation} \label{IIIlargeN}
B=-\frac{(2-\delta_{s,0})}{2^n} \, n! \, \lambda^{\frac{n}{2}}e^{\frac{y^2}{2}}\sum_{\delta=0}^{n}(2y)^{\delta}\sum_{l=0}^{\epsilon}(-1)^l\frac{N^{-l}}{\left(\epsilon-l \right)! \, \rho !}L_{N-s-1}^{s+l+1}(-4N y^2)
\end{equation}
and so
\begin{equation} \label{W1LagN-1}
\left< \trK e^a \, \tr a^n \right>_c = \left( \frac{\sqrt{\lambda}}{4N}\right)^n \frac{\dd^n}{\dd y^n} \left( L_{N-1}^1(-4N y^2) e^{2N y^2}\right)-
\end{equation}
$$
-\frac{(2-\delta_{s,0})}{2^n} \, n! \, \lambda^{\frac{n}{2}}e^{\frac{y^2}{2}}\sum_{\delta=0}^{n}(2y)^{s}\sum_{l=0}^{\epsilon}(-1)^l\frac{N^{-l}}{\left(\epsilon-l \right)! \, \rho!}L_{N-s-1}^{s+l+1}(-4N y^2) \, .
$$
For the large order Laguerre polynomial with negative argument there is an asymptotic form \cite{Laguerre}
\begin{equation}
L_{N-s-1}^{l+s+1}(- \nu \tilde{y}^2) e^{\frac{\nu \tilde{y}^2}{2}}=\frac{\alpha}{2^{l+s+1} F(\tilde{y})^{l+s+1}}I_{l+s+1}(\nu F(\tilde{y}))
\end{equation}
$$
\nu=4N+2l-2s \, , \,\, \tilde{y}=\frac{y}{\sqrt{1+\frac{l-s}{2N}}},
$$
$$
\alpha=\left( \frac{F(\tilde{y})}{\tilde{y}} \right)^{l+s+2}\left( \frac{\tilde{y}^2}{1+\tilde{y}^2} \right)^{\frac{1}{4}} \frac{1}{\sqrt{F(\tilde{y})}}=\alpha_0(\tilde{y})\left( \frac{F(\tilde{y})}{\tilde{y}} \right)^{l+s}
$$
And since the argument of the Bessel function is large, the asymptotic of Laguerre polynomial can be written as
\begin{equation} \label{as}
L_{N-s-1}^{l+s+1}(- \nu \tilde{y}^2) e^{\frac{\nu \tilde{y}^2}{2}}=\alpha_0(\tilde{y})\left( \frac{F(\tilde{y})}{\tilde{y}} \right)^{l+s}\frac{1}{2^{l+s+1} F(\tilde{y})^{l+s+1}}\frac{e^{\nu F(\tilde{y})}}{\sqrt{2 \pi \nu F(\tilde{y})}}
\end{equation}
Further simplification can be done if $K/N$ tends to a non-zero value. In this case of finite $K/N$ the differential operator in the first term of (\ref{W1LagN-1}) acts only on the exponent in the asymptotic representation (\ref{as}). Otherwise one needs also to take derivatives of other terms since they give terms proportional to $N/K$.
\par Using (\ref{WKlargeN}) one gets
\begin{equation} \label{SumBigN}
\frac{1}{N}\frac{\left< \trK e^a \, \tr a^n \right>_c}{\left< \trK e^a\right>} 
\approx \lambda^{\frac{n}{2}}\left(1+y^2 \right)^\frac{n}{2}-\frac{\lambda^{\frac{n}{2}}}{2^n}\frac{n!}{\left( \frac{n}{2}\right)!\left( \frac{n}{2}\right)!}-n! \, \frac{\lambda^{\frac{n}{2}}}{2^{n-1}}\sum_{p=1}^{\frac{n}{2}}\frac{\left(y-\sqrt{y^2+1} \right)^{2p}}{\left( \frac{n}{2}-p\right)!\left( \frac{n}{2}+p\right)!} \, .
\end{equation}
In order to get rid of the operator mixing appearing on $S^4$ one should use the following relation \cite{Sys}
\begin{equation} \label{rec}
\left< \trK e^a \, :\tr a^n : \right>_c=\sum_{l=0}^{\frac{n}{2}-1}(-1)^l A^n_l \left(\frac{\lambda}{4}\right)^l \left< \trK e^a \, \tr a^{n-2l} \right>_c \, ,
\end{equation}
where
\begin{equation}
A^n_l=\frac{n(n-l-1)!}{l!(n-2l)!}
\label{appendixc}\end{equation}
With (\ref{SumBigN}) and (\ref{rec}) one finds
\begin{equation} \label{W1KN}
\frac{1}{N}\frac{\left< \trK e^a \, :\tr a^n : \right>_c}{\left< \trK e^a \right>} \approx 2^{1-n}\sinh(n \arcsinh(y)) \, .
\end{equation}
which is in the agreement with \cite{Giombi} up to a normalization and coincides with \cite{Sys} exactly.

\subsection{Large N, $K/N \rightarrow 0$  limit}

Although it's difficult to find  the $N \rightarrow \infty, \, K/N  \rightarrow 0$ limit of (\ref{corK}) for general $n$, one can always check for particular values of $n$, that this limit of (\ref{corK}) is
\begin{equation} \label{W1K/N0}
\frac{1}{N}\frac{\left< \trK e^a \, :\tr a^n : \right>_c}{\left< \trK e^a \right>}=\frac{n}{2^n}\lambda^{\frac{n}{2}}I_n(K\sqrt{\lambda})
\end{equation}
which is in agreement with \cite{Sys}.
\par It's worth noting that unlike the case of finite $K/N$, the highest order in the large-$N$ expansion of (\ref{corK}) vanishes if $K/N \rightarrow 0$.
\par It also should be noted that (\ref{W1K/N0}) can not be found as the $K/N \rightarrow 0$ limit of (\ref{W1KN}) since in the derivation of (\ref{W1KN}) it was essential that $K/N$ was strictly positive, but can be found as the $K/N \rightarrow 0$ limit of (\ref{W1LagN-1}) with the help of the relation (\ref{rec}).

\section{Correlators of two Wilson loops}
In this section we consider the correlator of two $\frac{1}{2}$-BPS Wilson loops preserving the same subgroup of supersymmetries, \textit{i.e.} taken over the same circle and sharing the orientation in the internal space. For such Wilson loops the correlator also can be written as a matrix model integral due to localization \cite{Pestun}.
\begin{equation}
W_{\bf R, \, \bf R'}=\langle \tr_{\bf R}\, e^{\cal C} \tr_{\bf R'}\, e^{\cal C} \rangle =\frac{1}{N}\frac{1}{Z}\int  d a \, \Delta(a) \,e^{-   \sum_u {a_u^2}   }\, \, \tr_{\bf R} e^{  g\, a } \, \tr_{\bf R'} e^{  g\, a } \, .
\end{equation}
Using representation theory \cite{Jhones} one can write that
\begin{equation}
\tr_{\bf R}\, e^{ g\, a } \tr_{\bf R'}\, e^{ g\, a } = \tr_{\bf R \otimes \bf R'}\, e^{ g\, a }= \sum_{\bf {R_i}} C_{\bf R \otimes \bf R', R_i} \tr_{\bf R_i}\, e^{ g\, a } \, ,
\end{equation}
where $\bf R_i$ stands for an irreducible components in $\bf R \otimes \bf R'$ and $ C_{\bf R \otimes \bf R', R_i}$ are the corresponding Clebsch-Gordan coefficients.
\par For example, for a product of two symmetric Wilson loops (assuming $K \geq K'$) one gets a sum of traces in the two-rows representations
\begin{equation}
\tr_{K \otimes K'}\, e^{ g\, a }= \sum_{i=0}^{K'} \tr_{\{ K+i,K'-i \} }\, e^{ g\, a } \, .
\end{equation}
So the correlator of two symmetric Wilson loops can be written as 
\begin{equation}
W_{K, K'} =\sum_{i=0}^{K'} W_{ \{ K+i, K'-i \} }
\end{equation}
and hence in the limit of large coupling constant can be found exactly in terms of $N$ with (\ref{WL2rows}).

\subsection{Large $N$ limit}

Let us now consider a correlator of a WL in the fundamental representation with a WL in the representation associated with a  Young tableau with several lines (number of lines $n \ll N$) of the same length $\vec{K}=\{K_1, \, ..., \,K_n \}, \, K_i=K$.
\begin{equation} \label{corKf}
W_{\vec{K}, \, f} =W_{ \{ K_1+1, \, K_2, \, ..., \,K_n\} }+W_{ \{ K_1, \, K_2, \, ..., \,K_n, \, 1 \} }
\end{equation}
Letting $K$ to be of order $N$ one can apply (\ref{WLsevrows}) and write
$$
\frac{W_{ \{ K_1+1, \, K_2, \, ..., \,K_n\} }}{W_{\vec{K}}} \sim \frac{ e^{4N F(y+\frac{\sqrt{\lambda}}{4 N}) } \prod_{u=2}^{n} e^{4N F(y)}}{\prod_{u=1}^{n} e^{4N F(y)}}=
$$
\begin{equation} \label{1term}
=e^{\sqrt{\lambda}\frac{d}{d \,y}F(y)}=e^{\sqrt{\lambda}\sqrt{1+y^2}}.
\end{equation}
As for the second term, it is clear that in the leading order of large $N$
\begin{equation} \label{2term}
\frac{W_{ \{ K_1, \, K_2, \, ..., \,K_n, \, 1 \}}}{W_{\vec{K}}} \approx \frac{W_{\vec{K}} W_{f}}{W_{\vec{K}}} = W_{f} \sim e^{\sqrt{\lambda}}.
\end{equation}
Comparing (\ref{1term}) and (\ref{2term}) we see that  (\ref{1term}) is the leading term in the sum (\ref{corKf}) for any $y>0$, so (\ref{2term}) should be omitted.
\begin{equation} \label{corCFT}
\frac{W_{\vec{K}, \, f}}{W_{\vec{K}}} \sim e^{\sqrt{\lambda}\sqrt{1+y^2}}
\end{equation}
\par A correlator between a WL in the fundamental representation and a WL in the representation $\bf R$ associated with a Young tableau with many rows (number of lines $n \sim N$) of the same length was found in \cite{Pando}
\begin{equation}
\frac{W_{\bf R, \, f}}{W_{\bf R}} \sim e^{\sqrt{\lambda}(y+\sqrt{\frac{n}{N}})}+e^{\sqrt{\lambda}\sqrt{1-\frac{n}{N}}}.
\end{equation}
In this case an of two terms can be the leading one for some values of the parameters.
\par In order to compare the correlator (\ref{corCFT}) with the corresponding quantity on the AdS side, one has to find it also in the theory with $SU(N)$ gauge group. In the theory with $SU(N)$ group of symmetries there is an additional factor of $(\det e^{g \, a})^{-\frac{| \bf R |}{N}}$ with $| {\bf R} |= \sum_{i=0}^{g} K_i n_i$. In the case of several lines in a Young tableau ($n \ll N$) $\frac{| {\bf R} |}{N} \approx 0$, so the correlator does not change.

\subsection{String in degenerated genus one background}

As recently discussed in \cite{Pando}, according to the AdS/CFT correspondence, the correlator of Wilson loops of the form ${W_{\bf R, \, f}}/{W_{\bf R}}$ can be computed in the large ’t Hooft coupling constant limit as the on-shell action of a fundamental string in the bubbling geometries arising due to the backreaction of a WL in the representation $\bf R$.
\par  The metric of the bubbling geometries is the one associated with $AdS_2$, $S^2$ and $S^4$ fibration over a 2-dimensional complex Riemann surface $\Sigma$. All the geometric functions and fluxes can be expressed in terms of two holomorphic functions $\mathcal{A}$, $\mathcal{B}$ defined on the Riemann surface $\Sigma$.
\par Following the notations and approach of \cite{Pando} we take $\Sigma$ as a torus described by coordinates $(z, \, \bar{z})$ with periods $2\omega_1$ and $2\omega_3$ and write that in the case of genus one background corresponding to a rectangular Young tableau on the CFT side the functions $\mathcal{A}$, $\mathcal{B}$ are the following
\begin{equation}
\mathcal{A}=i \kappa_1 \left(\zeta(z-1)+\zeta(z+1)-2\frac{\zeta(\omega_3)}{\omega_3}z\right),
\end{equation}
$$
\mathcal{B}=i \kappa_2 \left(\zeta(z-1)-\zeta(z+1) \right),
$$
where $\zeta$ stands for the Weierstrass $\zeta$-function, a primitive of the Weierstrass $\wp$-function
\begin{equation}
\wp(z)=-\zeta'(z)
\end{equation}
and $\kappa_1$, $ \kappa_2$ are the constants defined by a requirement that the geometry reduces asymptotically to $AdS_5 \times S^5$ as
\begin{equation}
\kappa_1=\frac{L^2}{8}e^{-\frac{\Phi_0}{2}}\left(\wp(2)+\frac{\zeta(\omega_3)}{\omega_3}\right)^{-\frac{1}{2}}, \, \, \kappa_2=\frac{L^2}{8}e^{\frac{\Phi_0}{2}}\left(\wp(2)+\frac{\zeta(\omega_3)}{\omega_3}\right)^{-\frac{1}{2}}
\end{equation}
with $L^2=2 \sqrt{\pi N} \alpha'$, $e^{\frac{\Phi_0}{2}}=\sqrt{g_s}$.
\par The functions $\zeta(z)$, $\wp$ depend on the periods of the torus, which in their turn are specified by the two branch points $\tilde{e}_1$, $\tilde{e}_2$. Introducing also $\tilde{e}_3=-(\tilde{e}_1+\tilde{e}_2)$ we write the half-periods $\omega_1$, $\omega_3$ as
\begin{equation}
\omega_1=\frac{K\left(\frac{\tilde{e}_2 -\tilde{e}_3}{\tilde{e}_1-\tilde{e}_3}\right)}{\sqrt{\tilde{e}_1-\tilde{e}_3}}, \,\, \omega_3=i\frac{K\left(\frac{\tilde{e}_1 -\tilde{e}_2}{\tilde{e}_1-\tilde{e}_3}\right)}{\sqrt{\tilde{e}_1-\tilde{e}_3}},
\end{equation}
where $K$ is the complete elliptic integral of the first kind. Let us also introduce
\begin{equation}
 \omega_2=\omega_1+\omega_3, \,\, \omega_0=0.
\end{equation}
\par The parameters of the Young tableau defining the representation of the WL causing the geometry are related to the half-periods by the equations
\begin{equation} \label{nKviaom}
n=\frac{N \omega_3}{2 \pi i} \left( 4\left(\zeta(1)-\frac{\zeta(\omega_3)}{\omega_3} \right) + \frac{\left(\wp(1)+\frac{\zeta(\omega_3)}{\omega_3} \right)\wp''(1)-\wp'(1)}{\left(\wp(2)+\frac{\zeta(\omega_3)}{\omega_3} \right)\wp'(1)}\right),
\end{equation}
$$
K=\frac{\sqrt{\pi}i}{\omega_3}\sqrt{\frac{N}{g_s}}\left(\wp(2)+\frac{\zeta(\omega_3)}{\omega_3} \right)^{-\frac{1}{2}},
$$
where $n$ stands for the number of rows and $K$ for the length of rows of the rectangular Young tableau.
\par The string configuration extremizing the string action is the one with the world sheet extending all along the $AdS_2$, sitting at an arbitrary point both on $S^2$ and $S^4$ and at the points $z=\omega_{\alpha}, \, \alpha =0, \, 1, \, 2, \, 3$ on the complex plane $\Sigma$.
\par The on-shell action is
\begin{equation} \label{Sonshell}
S_{on-shell}(\omega_{\alpha})=-\frac{1}{\alpha'}\frac{L^2 \sqrt{g_s}}{4 \sqrt{\wp(2)+\frac{\zeta(\omega_3)}{\omega_3}}} \Bigg( 2 \zeta(2)-2[\zeta(1+\omega_{\alpha})+\zeta(1-\omega_{\alpha})]-\frac{\wp'(2)}{\wp(2)+\frac{\zeta(\omega_3)}{\omega_3}}+
\end{equation}
$$
+\Bigg| \frac{3\wp'(1+\omega_{\alpha})\left(\wp(1+\omega_{\alpha})+\frac{\zeta(\omega_3)}{\omega_3}\right)}{\wp''(1+\omega_{\alpha})-3\wp(1+\omega_{\alpha})\left(\wp(1+\omega_{\alpha})+\frac{\zeta(\omega_3)}{\omega_3}\right)} \Bigg|-\frac{3\wp'(1+\omega_{\alpha})\left(\wp(1+\omega_{\alpha})+\frac{\zeta(\omega_3)}{\omega_3}\right)}{\wp''(1+\omega_{\alpha})-3\wp(1+\omega_{\alpha})\left(\wp(1+\omega_{\alpha})+\frac{\zeta(\omega_3)}{\omega_3}\right)} \Bigg).
$$
To compare the minimalized string action with the correlator on the CFT side one has to express $\omega_{\alpha}$ through the parameters of the Young tableau $n, \,K$ with the help of equations (\ref{nKviaom}) and to substitute them in (\ref{Sonshell}). The relations (\ref{nKviaom}) were inverted in \cite{Pando} for the Young tableau with the number of lines $n$ being of order of $N$ and the length of the lines $K$ being of order of $N$ or larger. Let us do it for a small number of lines ($n \ll N$).
\par In this case the interval $[\tilde{e_1}\, \tilde{e_2}]$ collapses and hence the $AdS_5 \times S^5$ geometry is recovered \cite{Hoker, Benichou}. The half-period $\omega_1$ tends to infinity $\omega_1 = \mathcal{O}\left(\ln(\tilde{e_1}-\tilde{e_2})\right) = \mathcal{O}\left(\ln(N)\right)$ and therefore the Weierstrass elliptic functions can be written as
\begin{equation}
\zeta(z)=-\frac{\pi^2}{12 \omega_3^2} \left(1 + \frac{3}{\sinh^2} \left( \frac{i \pi z}{2 \omega_3}\right) \right) + \mathcal{O}\left( e^{-\frac{2 \pi i \omega_1}{\omega_3}}\right),
\end{equation}
$$
\wp = \frac{\pi^2 z}{12 \omega_3^2}+\frac{i \pi}{2 \omega_3} \coth\left(\frac{i \pi z}{2 \omega_3} \right)+ \mathcal{O}\left( e^{-\frac{2 \pi i \omega_1}{\omega_3}}\right).
$$
The half-period $\omega_3$ can be found from the second equation of (\ref{nKviaom}).
\begin{equation} \label{om3}
\omega_3= \frac{i \pi}{\sinh^{-1}(y)}, \,\,\, y=\frac{K \sqrt{\lambda}}{4 N}.
\end{equation}
Substituting (\ref{om3}) and infinite $\omega_1$ in (\ref{Sonshell}) one will find
\begin{equation} \label{Ssadp}
S(0)=S(\omega_3)=-\sqrt{\lambda}+\mathcal{O} \left(\frac{1}{N} \right),
\end{equation}
$$
S(\omega_1)=S(\omega_2)=-\sqrt{\lambda}\sqrt{1+y^2}+\mathcal{O}\left(\frac{1}{\sqrt{N}} \right).
$$
Again, as in the previous subsection, the first contribution to the action is suppressed comparing to the second one for any $y>0$. Hence we get in the leading order of the large $N$ expansion
\begin{equation}
e^{- S_{on-shell}} \approx e^{\sqrt{\lambda}\sqrt{1+y^2}},
\end{equation}
which coincide exactly with (\ref{corCFT}).

\acknowledgments
I would like to thank F.Fucito and J.F.Morales for having suggested the problem and for constant encouragment and advise during the completion of this work.
\par This work  is partially supported by the MIUR PRIN Contract 2015MP2CX4 ``Non-perturbative Aspects Of Gauge Theories And Strings''.

\end{document}